\documentclass[preprint2]{aastex}
\usepackage{graphicx}
\begin{document}
\title{Effects of Preheated Clusters on the CMB Spectrum}
\author{Kai-Yang Lin$^1$, Lihwai Lin$^1$, Tak-Pong Woo$^1$, 
Yao-Huan Tseng$^1$ 
\& Tzihong Chiueh$^{1,2}$}
\affil{1. Department of Physics, National Taiwan University,
Taipei, Taiwan}
\affil{2. Institute of Astronomy and Astrophysics, Academia Sinica, 
Taipei, Taiwan}

\begin{abstract}
Mounting evidence from $x$-ray observations reveals that bound objects 
should receive some form of energy in the past injected from
non-gravitaional sources.  We report that
an instantaneous heating scheme, for which
gases in dense regions were subjected to a temperature jump of $1$keV
at $z=2$ whereas those in rarified regions remained intact,
can produce bound objects obeying the observed mass-temperature
and luminosity-temperature relations.
Such preheating lowers the peak Sunyaev-Zeldovich (SZ) power
by a factor of $2$ and exacerbates
the need for the normalization of matter fluctuations $\sigma_8$
to assume an extreme high value $(\sim 1.1)$ for the SZ signals
to account for the excess anisotropy on 5-arcminute scale detected by the
Cosmic Background Imager in the cosmic microwave background radiation. 
   
\end{abstract}


\section{Introduction}

In the hierarchical structure formation model of cold dark matter,
gravitationally bound
objects ought to obey self-similar scaling relations. That is, 
at any cosmic epoch the mass of a collapse object scales as
the third power of its size,
and the squared velocity dispersion or temperature
scales as the second power of the size .  These scaling relations
follow from a straightforward dimensional analysis,
given that gravitational collapse occurs much faster than Hubble
expansion and depends on only two dimensional (quasi-)constants:
the gravitational constant $G$ and the background
matter density $\rho_{m0}$.  Although $\rho_{m0}$ varies on the Hubble
time scale, such dependence on long time scale affects only the interior
profiles of bound objects (Navarro et al. 1996; Shapiro \& Iliev
2002; Chiueh \& He 2002). 

However, $x$-ray observations in the past years
have consistently indicated that gases in bound objects violate
the self-similar scaling relations.
The observed $x$-ray luminosity-temperature ($L-T$)
relation shows that at a given luminosity the cluster temperature
appears considerably higher than the scaling relation ($L\propto T^2$),
more so for low-mass clusters than for high-mass clusters
(Ponman et al. 1996; Allen \& Fabian 1998; Xue \& Wu 2000).  
The observed mass-temperature (M-T)
relation also follows the same trend in violating the $M\propto T^{3/2}$
scaling law (Finoguenov et al. 2001).  
More remarkably, recent XMM observations of metal lines that
directly probe the core temperatures of cooling-flow clusters find that
gases of temperature below some fraction of bulk temperature are
unexpectedly rare (Peterson et al. 2002), suggestive of an  
entropy floor at the cluster core (Ponman,
Cannon \& Navarro 1999; Loewenstein 2000) thus requiring
some form of non-gravitational energy injection to compensate for
Bremsstrahlung cooling.

One candidate for the quested heat sources in clusters and groups is
the outflow from Active Galactic Nucleus (AGN).
(Tabor \& Binney, 1993; David et al. 2001).
If this is correct, such a mechanism should be more
pronounced in the far past when quasars were highly active 
than the
present.  For this and other reasons, investigators have tested
various schemes for heat injection at $z \geq 4$ (da Silva et
al. 2001; Bialek et al. 2001; Muanwong et al. 2002).  
However, recent combined analyses of the ASCA data on the entropy
profiles in galaxy groups and the structure formation model
indicate that preheating should occur impulsively around $z=2 - 2.5$ 
(Finoguenov et al. 2002).   A similar conclusion has also been 
reached from a different perspective, using
the constraint on Compton $y$ ($<1.5\times 10^{-5}$) 
provided by the COBE/FIRAS 
data in testing the SZ effect at the hot spots of
AGN jets (Yamada \& Fujita 2001).  The present work has been so motivated
to consider impulsive heating around $z\sim 2$.  

On the other hand, the recent Cosmic Background Imager (CBI) experiment
discovered that the temperature anisotropy of cosmic-microwave-background
radiation (CMB) exhibits a $3\sigma$ higher level of anisotropy,
around the harmonic mode number
$2000< l < 3500$, than anticipated with Silk damping 
(Pearson et al. 2002).  This
feature has been attributed to the large SZ flux (Bond et al. 2002).  
In order to account for
such a large SZ flux, the value of
$\sigma_8$ needs to be at least as large as unity, though from
Figs.10 and 11 of Bond et al.(2002), a slightly larger 
$\sigma_8$ fits the CBI data even better.  

The SZ flux has been known to depend very sensitively on
$\sigma_8$ (Fan \& Chiueh 2001; Komatsu \& Seljak 2002), due
to the sensitive dependence of cluster number density on the amplitude
of matter fluctuations
at the high end of mass function.  Hence the SZ effect can serve 
as a unique probe to fix $\sigma_8$ to high precision.
As of today, the large-scale galaxy distribution combined with
distant supernova and CMB data have constrained $\sigma_8$ 
within a narrow window $0.75 <\sigma_8< 1$ (Bond et al. 2002).  
From the perspective of 
cluster abundance explored by the Sloan
Digital Sky Survey (SDSS) (Bahcall et al. 2002),
the extreme value $\sigma_8=1$
has, however, placed a considerably low ceiling for the matter
density $\Omega_m = 0.175\pm 0.025$.

If the SZ effect indeed accounts for the CBI deep-field excess,
preheating can actually exacerbate
the awkward situation pertinent to a high value of $\sigma_8$.
Past works have indicated that 
preheating generally reduces the SZ flux (Bialeck et al, 2001; da Silve
et al. 2001).  However, these works were not able to reproduce
both observed $L-T$ and $M-T$ relations and hence the reduction 
factor for SZ flux cannot be reliably gauged.
The present paper reports that impulsive heating around $z=2$ can 
reproduce the
observed $L-T$ and $M-T$ relations and yields a modified SZ 
power spectrum lower than that without preheating by a factor $2$. 
It therefore pushes
$\sigma_8$ upward by a factor $1.1$ for SZ effects to explain the CBI
high-$l$ result, and $\Omega_m$ downward by $1.2$ constrained by 
local cluster abundance. 
\section{Simulations, M-T and L-T Relations}
Cosmological hydrodynamical simulations with the GADGET code
(Springel, Yoshida \& White 2001) were run 
for both $\sigma_8=0.94$ and $1$ cases in a periodic 
box of comoving size $100 h^{-1}$ Mpc on our 32-node,
dual-cpu Athlon cluster.  The simulations contain
$256^3/2$ dark matter particles and $256^3/2$
gas particles, and the gravitational softening length is
chosen to be $20 h^{-1}$ kpc. 
Both adiabatic and preheating runs start at $z=100$
with the same initial conditions of $n_s=1$ and
$(\sigma_8, \Omega_m, \Omega_b, \Omega_\Lambda, h)
=(0.94, 0.34, 0.05, 0.66, 0.66)$ and 
$(1, 0.3, 0.044, 0.7, 0.64)$ separately.

Preheating takes place at $z=2$ with a density-dependent 
injection energy per mass,
\begin{equation}
\Delta u=u_0 \exp[-\frac{\beta\rho_{m0}}{\rho_m}],
\end{equation}
where $\rho_{m0}$ is the background matter density. Both $u_0$ and
$\beta$ are parameters to be tuned to fit the observed
$M-T$ and $L-T$ relations described below.  This form of energy injection
ensures that the low-density regions receive no heat, giving rooms
for some cool gases to survive and
the high density regions receive a constant energy per mass $u_0$.
The injected entropy has minima at the cores of bound objects 
and maxima at the outskirts around $\rho_m\sim \beta\rho_{m0}$.
This heating scheme is similar, though not in detail, 
to the entropy injection of Borgani et al.(2001 \& 2002).
The best parameters are found to be $\beta=1$ and $u_0=(500$
km/sec)$^2$.  Such a value of $u_0$
corresponds to $1$ keV temperature, and the amount of
injected heat is equivalent to giving every baryon
a $0.5$ keV temperature jump.  This amount is roughly 
consistent with those suggested by recent studies on
heat injection from the bound-object interiors
(Loewenstein 2000; Wu, Fabian \& Nulsen
2000; Bower et al. 2001; Borgani et al. 2001 \& 2002; Xue \& Wu 2002).


Standard FOF algorithm
with a $0.2$ linking length is employed to locate halos.
The mass centers may considerably deviate from the peaks 
of spherical over-density, such as in the case of major mergers.
In these cases, the over-density so obtained is not useful for
testing the $L-T$ and $M-T$ relations.  They are
avoided by selecting only those haloes with nonzero
$M_{2500}$, where $M_{n}$ represents the mass within a sphere of
over-density $n$ times that of the critical density.
The temperature $T$ of an ionized plasma
is related to the internal energy per mass $u$ of the neutral-fluid in
hydrodynamic simulations by $k_B T=(2/3)\mu m_p u$,
where $\mu$ is the mean "molecular" weight of $e^-$, $p^+$ and
$He^{2+}$ and equal to $0.588$.  In addition,
we define the Bremsstrahlung emission-weighted temperature
$\langle n_e^2 T^{3/2}\rangle/\langle n_e^2 T^{1/2}\rangle$,
to be compared with the measured temperature of $x$-ray continuum.

Figure 1 compares both adiabatic and preheating 
$M_{500}$-$T_{500}$ relations from our $\sigma_8=0.94$ and $1$
simulations with the fitting formula given by
Evrard, Metzler \& Navarro (1996) from their adiabatic simulations, 
and with the fitting formula of
Finoguenov, Reiprich, \& B\"{o}hringer (2001) derived
from the observational data re-scaled to $z=0$. 
Only data
with luminosity higher than $10^{41}$ erg/sec are plotted here.
The $M_{500}-T_{500}$ relation of the adiabatic simulation agrees fairly
with that of Evrard et al., but both predict too low a
temperature in bound objects.  On the other hand, our preheating
simulations yield $M_{500}-T_{500}$ relations of $\sigma_8=0.94$ and $1$
agreeing
quite well with the observations.  
\begin{figure}[t]
 \plotone{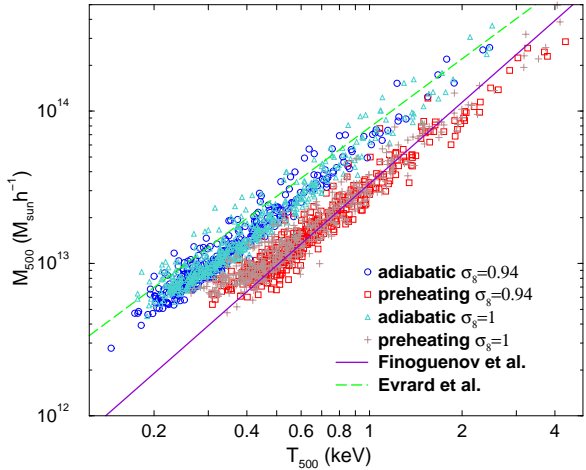}
 \caption{$M_{500}-T_{500}$ relation: Open circles (triangles) represent
adiabatic 
  simulation data for $\sigma_8=0.94 (1)$,
  to be compared with the long-dashed line, 
 the simulation results of
 Evrard et al.(1996). Open squares (crosses) are
 preheating simulation data for $\sigma_8=0.94 (1)$ and are consistent
 with the solid
 line, the observational result given in Finoguenov et al.(2001). 
 All results are for $z\sim 0$.}
 \label{fig.M-T}
\end{figure}

The $x$-ray luminosity 
requires somewhat sophisticated treatments
(Mathiesen \& Evrard 2001), since line emission below $0.5$ keV can
vastly dominate the Bremsstrahlung continuum.
A constant metalicity, $0.3$ times the solar
value, is adopted and the bolometric luminosity reads
\begin{equation}
L_{500}=1.4\times10^{-27}\int T^{1/2}g(T)
n_e(\sum_i n_i Z_i^2)dV
\end{equation}
in unit of erg/sec,
where the volume integration is up to $R_{500}$, the summation
is over different ion species, $Z$ is the atomic number,
and $g(T)$ the Gaunt factor.  Figure 2 shows $L_{500}$ 
against the emission weighted $T_{500}$.  Also shown are
the observed $L_{500}$ and $T_{500}$ compiled from Arnaud \& Evrard
(1999), Helsdon \& Ponman (2000) and Markevitch (1998).  The preheating
results for both $\sigma_8=0.94$ and $1$ cases again show excellent
agreement with $x$-ray observations.
Interestingly, Novicki, Sornig \& Henry (2002) pointed out that
the $L-T$ relation is consistent with no
evolution in a low-density universe.  Our preheating
result does display this tendency within $0<z<1$.  This behavior 
also suggests that after initial transients, the heated clusters quickly
assume certain relaxed states since $z=1$.

\begin{figure}[t]
 \plotone{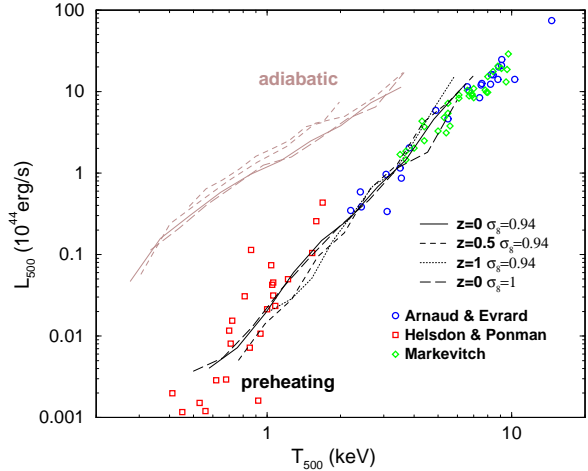}
 \caption{$L_{500}-T_{500}$ relation: square, circle, and diamond symbols
are 
 the observed x-ray
 luminosities and temperatures given by different groups as described
 in the text. The solid, dashed, dotted and long-dashed lines
 are the simulation results for $\sigma_8=0.94$ at $z=$0, 0.5 and 1,
 and for $\sigma_8=1$ at $z=0$,
 respectively. The
 upper band is for the adiabatic simulations, and the lower band
 for the preheating simulations. Both display little evolution
 within $0<z<1$.}
 \label{fig.L-T}
\end{figure}

\section{SZ and CMB Power Spectra}

The agreement of our $M-T$ and $L-T$ relations with observations
confirms the proposed heating scheme to be quantitatively viable.
We thus proceed to construct the $1$ deg$^2$
SZ map by projecting the electron pressure through 
the randomly displaced and oriented simulation 
boxes, separated by $100 h^{-1}$Mpc, along a viewing cone 
to the redshift 
direction up to $z=2.5$.  The average power spectra
of $40$ such SZ maps for each adiabatic and preheating
simulation of $\sigma_8=0.94$ and $1$ cases are shown in Fig.3. 
The adiabatic results are consistent with earlier GADGET results
(Springel et al. 2001), where small-scale structures
are more abundant than those in grid-based hydrodynamic simulations
(Zhang et al. 2002).
Both $\sigma_8=0.94$ and $1$ runs show that preheating lowers
the averaged SZ power spectrum by a factor $2$ on few-arcminute 
scales.  

In Fig.3 we have rescaled the power spectra of the
$\sigma_8=1$ case from original $\Omega_b h^2=0.018$
to $\Omega_b h^2=0.022$ so as to be compared with those analyzed 
in Bond et al.(2002).  The CBI deep-field observations 
scanned a 9 deg$^2$ sky (Pearson et al. 2002).
We hence compute the {\it preheated} SZ power spectrum
averaged over 9 one-deg$^2$ SZ maps, which is then added to
the (error-free) primary CMB spectrum of
$\Omega_b=0.04, \Omega_m=0.3, \Omega_\Lambda=0.7, n_s=0.975,
h=0.68$, parameters used in Fig.11 of Bond et al.(2002), to obtain one
realization. A total of 40 realizations are constructed.  

The preheated SZ power spectrum of $\sigma_8=1$ looks very much
like the adiabatic power spectrum of $\sigma_8=0.9$, according to
the scaling law given by 
Komatsu \& Seljak (2001) that the peak SZ power approximately 
scales as $\sigma_8^7$.
The $1\sigma$ error bars for the $\sigma_8=1$
preheated SZ$+$CMB power arise from the cosmic variance of SZ clusters 
and are seen to be within a factr $1.3$ from the mean.
Also clear in Fig. 3 is that the preheated
SZ$+$CMB power spectrum nearly misses the 2-$\sigma$ error
box of band-averaged power in CBI deep data by $1\sigma$.  
Such reduction of SZ power
by preheating makes it implausible for the SZ effect 
to explain the excess flux detected by CBI around $2000<l<4000$
under the framework of concordance model with $\sigma_8=1$. 
Employing the $\sigma_8^7$ scaling law,
we estimate that the required 
$\sigma_8$ needs to be raised to $1.1$ in order to be
consistent with the CBI high-$l$ excess. 

\begin{figure}[t]
 \plotone{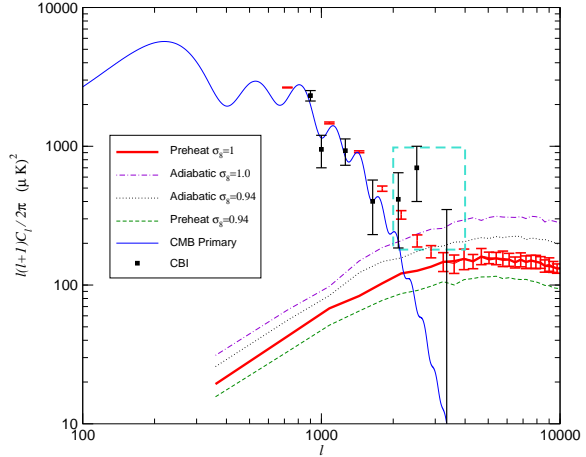}
 \caption{Comparison of CBI high-$l$ data with the averaged
preheating and adibatic SZ power spectra for $\sigma_8=0.94 \& 1$,
and with the CMB$+(\sigma_8=1$) preheating SZ spectrum.  
The $1\sigma$ error bars of the latter arise from
the cosmic variance of SZ spectra over 40 fields of 9
deg$^2$ and nearly miss the $2\sigma$ error box of
the band-averaged power in CBI deep data.}
\label{fig.3}
\end{figure}


\section{Conclusions}

The agreement of our preheating cosmological simulations with
the observed $M-T$ and $L-T$ relations 
demonstrates that the impulsive 
heating scheme described by Eq.(1) is a plausible 
prescription, though it is probably not the unique one and
its physical origin is unclear.
Contrary to our
results, recent simulations of Borgani et al.(2002) adopting 
entropy injection failed in
producing the observed $M-T$ relation despite their successful
reproduction of the $L-T$ relation.
Various preheating schemes so far reported tend to yield
reduced SZ flux, which in our case
can be attributed to that gases 
impulsively blown away from the gravitational potential 
well form an extended halo, which can subsequently not fall into,
therefore compressionally heated by, the
ever-growing potential well.  The agreement of our results 
with the observed $M-T$ and $L-T$ relations encourages us  
to believe that our quantitative prediction of the preheated SZ flux
reduced by a factor $2$ is founded on an empirically sound base.

This reduction of SZ flux has a significant implication to the
interpretation of the CBI deep-field results, in that the measured
$3\sigma$ enhancement of CMB high-$l$ anisotropy is unlikely to be caused
by the SZ effect of intra-cluster gases in a $\sigma_8=1$ universe
of concordance model
proposed in the original CBI paper (Bond et al. 2002). 
We estimate that $\sigma_8$ needs to assume a value
$1.1$ for this mechanism to work. 
The unfortunate consequence of it is that $\Omega_m$ is then pushed to 
lie about $0.15\pm 0.02$, 
constrained by $\sigma_8\Omega_m^{0.6}=0.35\pm 0.03$
of recent SDSS cluster abundance studies (Bahcall et al. 2002).
Such an $\Omega_m$ can be just too low to be consistent with
the observed $x$-ray baryon fraction $(\leq 1/4)$ in massive clusters
(Ettori \& Fabian 1999; Mohr et al. 1999).


\acknowledgements

We thank M. Birkinshaw, C.J. Ma, U. Pen, K. Umetsu, 
D. Worrall, P.J.H. Wu, X.P. Wu, Y.J. Xue and 
P. Zhang for various helpful discussions. Supports from NSC of Taiwan, 
under the grant NSC90-2112-M-002-026, and uses of the NCHC computing
facilities are acknowledged.

\end{document}